%% file: tmparxiv.tex
\documentclass[1opt,twocolumn,letter]{IEEEtran}

\usepackage[pdftex]{graphicx}
\DeclareGraphicsExtensions{.pdf}
\graphicspath{{fig/}}
\usepackage[tight]{subfigure}
\usepackage{fancyhdr}
\usepackage{amsmath}
\usepackage[ruled,slide, vlined,linesnumbered]{algorithm2e}
\usepackage{tikz}
\usepackage[sc, font=it]{caption}

\makeatletter \newcommand \listoftodos{\section*{Todo list} \@starttoc{tdo}}
\newcommand\l@todo[2]{
  \par\noindent \textit{#2}, \parbox{10cm}{#1}\par
} 
\makeatother

\newcommand{\ignore}[1]{{}}

\begin{document}

\title{Fault Tolerant QR Factorization\\
  for General Matrices}

\author{
\begin{tabular}{cc}
Camille Coti\\
\end{tabular}\\
\textit{LIPN, CNRS, UMR 7030} \\
\textit{Universit\'e Paris 13, Sorbonne Paris Cit\'e}\\
\textit{F-93430, Villetaneuse, France}\\
\textit{camille.coti@univ-paris13.fr}
}

\maketitle

\thispagestyle{empty}

\begin{abstract}
This paper presents a fault-tolerant algorithm for the QR factorization of
general matrices. It relies on the communication-avoiding algorithm, and uses
the structure of the reduction of each part of the computation to introduce
redundancies that are sufficient to recover the state of a failed
process. After a process has failed, its state can be recovered based on the
data held by one process only. Besides, it does not add any significant
operation in the critical path during failure-free execution. 
\end{abstract}



\section{Introduction}
\label{sec:intro}

Fault tolerance for high performance distributed applications can be
achieved at system-level or application-level. System-level fault tolerance is 
transparent for the application and requires a specific middleware that can
restart the failed processes and ensure coherent state of the application
\cite{fgcs08,BLKC04}. 

Application-level fault tolerance requires the application itself to
handle the failures and adapt to them. Of course, it implies that the
middleware that supports the distributed execution must be robust
enough to survive the failures and provide the application with
primitives to handle them \cite{FTMPI}. Moreover, it requires that
the application uses fault-tolerant algorithms that can deal with
process failures \cite{BDDL09}.

Recent efforts in the MPI-3 standardization process \cite{mpi3}
defined an interface for a mechanism called \emph{User-Level Failure
  Mitigation} (ULFM) \cite{BBHHBD13} and \emph{Run-Through
  Stabilization} \cite{HGBBPS11}.

This paper deals with the QR factorization of general
matrices. After a quick overview of techniques for fault tolerance
(section \ref{sec:abft}), we describe the communication-avoiding
QR factorization algorithm we are relying on in this paper in section
\ref{sec:caqr:algo}. Then we give the full fault-tolerant algorithm in sections
\ref{sec:caqr:fttsqr} for the panel and \ref{sec:caqr:ftcaqr} for the trailing
matrix. 

\section{Algorithm-Based Fault Tolerance}
\label{sec:abft}

FT-MPI~\cite{FTMPI, FGBACPLD04} defined four error-handling semantics
that can be defined on a communicator. \emph{SHRINK} consists in
reducing the size of the communicator in order to leave no hole in it
after a process of this communicator died. As a consequence, if one
process $p$ which is part of a communicator of size $N$ dies, after
the failure the communicator has $N-1$ processes numbered in
$[0,N-2]$. On the opposite, \emph{BLANK} leaves a hole in the
communicator: the rank of the dead process is considered as invalid
(communications return that the destination rank is invalid), and
surviving processes keep their original ranks in $[0,N-1]$. While
these two semantics survive failures with a reduced number of
processes, \emph{REBUILD} spawns a new process to replace the dead
one, giving it the place of the dead  process in the communicators it
was part of, including giving it the rank of the dead process. Last,
the \emph{ABORT} semantics corresponds to the usual behavior of
non-fault-tolerant applications: the surviving processes are
terminated and the application exits.

Using the first three semantics, programmers can integrate
 failure-recovery strategies directly as part of the algorithm that
 performs the computation. For instance, diskless
 checkpointing \cite{PLP98} uses the memory of other processes to save
 the state of each process. Arithmetic on the state of the processes
 can be used to store the checksum of a set of
 processes \cite{CFGLABD05}. When a process fails, its state can be
 recovered from the checkpoint and the states of the surviving
 processes. This approach is particularly interesting for iterative
 processes. Some matrix operations exhibit some properties on this
 checkpoint, such as \emph{checkpoint invariant} for LU
 factorization \cite{DBBHD12}. 

A proposal for \emph{run-through stabilization} introduced new
constructs to handle failures at communicator-level 
\cite{HGBBPS11}. Other mechanisms, at process-level, have been
integrated as a proposal in the MPI 3.1 standard draft \cite[ch
15]{mpi-3.1}. It is called \emph{user-level failure
mitigation} \cite{BBHHBD13}. Failures are detected when an operation
involving a failed process fails and returns an error. As a
consequence, operations that do not involve any failed process can
proceed unknowingly.  

\section{Fault-tolerant communication-avoiding QR factorization}
\label{sec:caqr}

In this section, we first recall how communication-avoiding QR works in section
\ref{sec:caqr:algo}. Then we give the fault-tolerant algorithm in two parts: for
the processes involved in the panel factorization in section
\ref{sec:caqr:fttsqr}, and for the processes involved in the update of the
trailing matrix in section \ref{sec:caqr:ftcaqr}.

\subsection{CAQR algorithm}
\label{sec:caqr:algo}

Communication-avoiding algorithms were introduced in \cite{CAQR}
\cite{CAQRLU}. They minimize the number of communications, at the cost
of some extra computations. Given the relative computation vs
communication speeds of the current architectures, these algorithms
are faster than traditional algorithms that maximize the parallelism
between the processing elements and involve more communications on a
wide range of architectures, from multicores \cite{DGG10} to grids
\cite{ACDHL10} and GPUs \cite{BDDGRT12}.

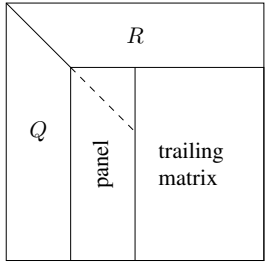
\begin{figure}[Hht]
\begin{center}
\resizebox{.4\linewidth}{!}{\input{fig_panel.tex}}
\caption{\label{fig:panelupdate}Panel/update organization of the QR factorization.}
\end{center}
\end{figure}

CAQR relies on two operations: a panel factorization and an update of
the trailing matrix. A set of columns on the left of the matrix is used as a
\emph{panel}. The panel is factorized and, using the result of the
factorization, the part of the matrix on the right of this panel, called the
\emph{trailing matrix}, is updated. This organization is represented in Figure
\ref{fig:panelupdate}.

The algorithm can be decomposed as follows on a
matrix A that can be represented by blocks:

  \begin{center}
  $ A = \begin{pmatrix} A_{11} & A_{12} \\
    A_{21} & A_{22} \end{pmatrix}
  = Q_1 \begin{pmatrix} R_{11} & R_{12} \\
    0 & A_{22}^1  \end{pmatrix} $
  \end{center}

  \begin{enumerate}
  \item Panel factorization:\label{step:panel}
    $ \begin{pmatrix} A_{11}  \\
    A_{21} \end{pmatrix}
    = Q_1 \begin{pmatrix} R_{11} \\
    0  \end{pmatrix} $
  \item Compact representation: 
    $Q_1 = I - Y_1 T_1 Y_1^T$
  \item Update the trailing matrix:\\
    $ \begin{pmatrix} I - Y_1 T_1 Y_1^T  \end{pmatrix}
    \begin{pmatrix} A_{12} \\ A_{22}\end{pmatrix}\\
    = \begin{pmatrix} A_{12} \\ A_{22}\end{pmatrix}
    - Y_1 ( T_1 ^ T ( Y_ 1 ^T \begin{pmatrix} A_{12}\\ A_{22}\end{pmatrix}))
    = \begin{pmatrix} R_{12} \\ A_{22} ^1 \end{pmatrix} $
  \item Continue recursively on the submatrix $A_{22} ^1$
\end{enumerate}
  
The panel factorization (step \ref{step:panel}) is a specific kind
of QR factorization. Since it factorizes a matrix with a particular
shape (called \emph{tall and skinny}), a dedicated algorithm is
used: TSQR \cite{BDGJNS14} \cite{L10}.

\subsection{Fault-tolerant TSQR}
\label{sec:caqr:fttsqr}

\begin{figure}[Hht]
\resizebox{\linewidth}{!}{
\begin{tikzpicture}
\huge


\foreach \y in {0, 1, 2, 3}{
  \path[draw,thick] ( 0, -4*\y ) rectangle ( 1.8, -4*\y-3.8 );
    \node at ( -1, -4*\y-.8 ) {$\mathbf{P_\y}$ };
    \node at ( 1, -4*\y-.8 ) {$A_\y$ };
}


\foreach \y in {0, 1, 2, 3}{
  \path[draw,thick] ( 4, -4*\y ) rectangle ( 5.8, -4*\y-3.8 );
    \node at ( 5.2, -4*\y-.6 ) {$R_\y$ };
    \node at ( 4.5, -4*\y-2.5 ) {$V_\y$ };
    \draw [ thick] ( 4, -4*\y ) -- ( 5.8, -4*\y-1.8 );
    \draw [ ->, thick ] (2, -4*\y-1) -- (3.8, -4*\y-1 );
}
  \node at ( 3, 1.5 ) {\textit{QR} };


  \foreach \y in {0, 2}{
  \path[draw,thick] ( 8, -4*\y ) rectangle ( 9.8, -4*\y-3.6 );
    \node at ( 9.2, -4*\y-.6 ) {$R_\y$ };
    \pgfmathsetmacro\z{\y + 1};
    \node at ( 9.2, -4*\y-2.4 ) {$R_{\pgfmathprintnumber{\z}}$ };
   \draw [thick] ( 8, -4*\y-1.8 ) -- ( 9.8, -4*\y-1.8 );
  \draw [ thick] ( 8, -4*\y ) -- ( 9.8, -4*\y-1.8 );
    \draw [thick] ( 8, -4*\y-1.8 ) -- ( 9.8, -4*\y-3.6 );
     \draw [  ->, thick ] (6, -4*\y-5) -- (7.8, -4*\y-2.8 );
   \draw [  ->, thick ] (6, -4*\y-1) -- (7.8, -4*\y-1 );
}
  \node at ( 7, 1.5 ) {\textit{Send/Recv} };


  \foreach \y in {1, 3}{
  \path[draw,thick, dashed] ( 8, -4*\y ) rectangle ( 9.8, -4*\y-3.6 );
    \pgfmathsetmacro\z{\y-1};
    \node at ( 9.2, -4*\y-.6 ) {$R_{\pgfmathprintnumber{\z}}$ };
    \pgfmathsetmacro\z{\y};
    \node at ( 9.2, -4*\y-2.4 ) {$R_{\pgfmathprintnumber{\z}}$ };
   \draw [thick, dashed] ( 8, -4*\y-1.8 ) -- ( 9.8, -4*\y-1.8 );
  \draw [ thick, dashed] ( 8, -4*\y ) -- ( 9.8, -4*\y-1.8 );
    \draw [thick, dashed] ( 8, -4*\y-1.8 ) -- ( 9.8, -4*\y-3.6 );
     \draw [  ->, thick, dashed ] (6, -4*\y+3) -- (7.8, -4*\y-.8 );
   \draw [  ->, thick, dashed ] (6, -4*\y-1) -- (7.8, -4*\y-3 );
}



  \foreach \y in {0, 2}{
  \path[draw,thick] ( 12, -4*\y ) rectangle ( 13.8, -4*\y-3.6 );
    \node at ( 13.2, -4*\y-.6 ) {$R_\y'$ };
    \node at ( 12.5, -4*\y-2.5 ) {${V_\y}'$ };
    \draw [ thick] ( 12, -4*\y ) -- ( 13.8, -4*\y-1.8 );
  \draw [  ->, thick ] (10, -4*\y-1) -- (11.8, -4*\y-1 );
}

  \foreach \y in {1, 3}{
  \path[draw,thick, dashed] ( 12, -4*\y ) rectangle ( 13.8, -4*\y-3.6 );
     \pgfmathsetmacro\z{\y-1};
   \node at ( 13.2, -4*\y-.6 ) {$R_{\pgfmathprintnumber{\z}}'$ };
    \node at ( 12.5, -4*\y-2.5 ) {${V_{\pgfmathprintnumber{\z}}}'$ };
    \draw [ thick, dashed] ( 12, -4*\y ) -- ( 13.8, -4*\y-1.8 );
  \draw [  ->, thick, dashed ] (10, -4*\y-1) -- (11.8, -4*\y-1 );
}

  \node at ( 11, 1.5 ) {\textit{QR} };
 

  \path[draw,thick] ( 16, 0 ) rectangle ( 17.8, -3.6 );
  \draw [ thick] ( 16,0 ) -- ( 17.8, -1.8 );
    \draw [ thick] ( 16, -1.8 ) -- ( 17.8, -3.6 );
    \draw [thick] ( 16, -1.8 ) -- ( 17.8, -1.8 );
    \node at ( 17.2, -.6 ) {$R_0'$ };
    \node at ( 17.2, -2.4 ) {$R_2'$ };
  \draw [  ->, thick ] (14, -1) -- (15.8, -1 );
  \draw [  ->, thick ] (14, -9) -- (15.8, -3 );


  \path[draw,thick,loosely dashed] ( 16, -8 ) rectangle ( 17.8, -11.6 );
  \draw [ thick, loosely dashed] ( 16,-8 ) -- ( 17.8, -9.8 );
    \draw [ thick, loosely dashed] ( 16, -9.8 ) -- ( 17.8, -11.6 );
    \draw [thick, loosely dashed] ( 16, -9.8 ) -- ( 17.8, -9.8 );
    \node at ( 17.2, -8.6 ) {$R_0'$ };
    \node at ( 17.2, -10.4 ) {$R_2'$ };
  \draw [  ->, thick,loosely  dashed ] (14, -1) -- (15.8, -9 );
 \draw [  ->, thick,loosely  dashed ] (14, -9) -- (15.8, -11 );


  \path[draw,thick,  dashed ] ( 16, -4 ) rectangle ( 17.8, -7.6 );
  \draw [ thick,  dashed] ( 16,-4 ) -- ( 17.8, -5.8 );
    \draw [ thick,  dashed] ( 16, -5.8 ) -- ( 17.8, -7.6 );
    \draw [thick,  dashed] ( 16, -5.8 ) -- ( 17.8, -5.8 );
    \node at ( 17.2, -4.6 ) {$R_0'$ };
    \node at ( 17.2, -6.4 ) {$R_2'$ };
  \draw [  ->, thick, dashed ] (14, -5) -- (15.8, -5 );
 \draw [  ->, thick, dashed ] (14, -13) -- (15.8, -7 );

  \path[draw,thick,  dashed ] ( 16, -12 ) rectangle ( 17.8, -15.6 );
  \draw [ thick,  dashed] ( 16,-12 ) -- ( 17.8, -13.8 );
    \draw [ thick,  dashed] ( 16, -13.8 ) -- ( 17.8, -15.6 );
    \draw [thick,  dashed] ( 16, -13.8 ) -- ( 17.8, -13.8 );
    \node at ( 17.2, -12.6 ) {$R_0'$ };
    \node at ( 17.2, -14.4 ) {$R_2'$ };
  \draw [  ->, thick, dashed ] (14, -5) -- (15.8, -13 );
 \draw [  ->, thick, dashed ] (14, -13) -- (15.8, -15 );

  \node at ( 15, 1.5 ) {\textit{Send/Recv} };


  \path[draw,thick] ( 20, 0 ) rectangle ( 21.8, -3.6 );
    \node at ( 21.2, -.6 ) {$R$ };
    \node at ( 20.5, -2.5 ) {$V$ };
  \draw [ ->, thick ] (18, -1) -- (19.8, -1 );
  \draw [ thick] ( 20,0 ) -- ( 21.8, -1.8 );

  \path[draw,thick, dashed] ( 20, -4 ) rectangle ( 21.8, -7.6 );
    \node at ( 21.2, -4.6 ) {$R$ };
    \node at ( 20.5, -6.5 ) {$V$ };
  \draw [ ->, thick,  dashed ] (18, -5) -- (19.8, -5 );
  \draw [ thick,  dashed] ( 20, -4 ) -- ( 21.8, -5.8 );

  \path[draw,thick, loosely dashed] ( 20, -8 ) rectangle ( 21.8, -11.6 );
    \node at ( 21.2, -8.6 ) {$R$ };
    \node at ( 20.5, -10.5 ) {$V$ };
  \draw [ ->, thick, loosely dashed ] (18, -9) -- (19.8, -9 );
  \draw [ thick, loosely dashed] ( 20, -8 ) -- ( 21.8, -9.8 );

  \path[draw,thick, dashed] ( 20, -12 ) rectangle ( 21.8, -15.6 );
    \node at ( 21.2, -12.6 ) {$R$ };
    \node at ( 20.5, -14.5 ) {$V$ };
  \draw [ ->, thick, dashed ] (18, -13) -- (19.8, -13 );
  \draw [ thick, dashed] ( 20, -12 ) -- ( 21.8, -13.8 );

  \node at ( 19, 1.5 ) {\textit{QR} };
\end{tikzpicture}
}
\captionof{figure}{\label{fig:rtsqr}Computing the R of a matrix using
a TSQR factorization on 4 processes with redundant $\widetilde{R}$ factors.}
\end{figure}
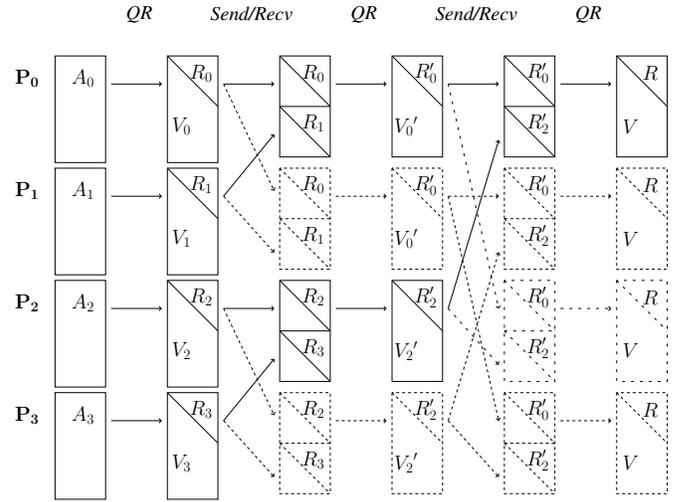

In \cite{C16}, we have presented a set of algorithms to achieve fault
tolerance in the TSQR panel factorization. The idea was to exploit the
idle processes along the reduction tree in order to integrate
redundancy with a very low overhead. Instead of just having odd-number
(modulo the step number) processes sending their intermediate $\tilde{R}$
factor to an even-numbered (modulo the step number) process and stop
computing, the two processes exchange their intermediate $\tilde{R}$
factors and both compute the same new intermediate $\tilde{R}$ factor. In
other words, the reduction turns into an all-reduce operation, where
the number of processes that own the same data (and therefore, the
resilience of the computation) doubles at each step (see Figure
\ref{fig:rtsqr}). 

This process has shown to have little overhead during fault-free execution and
potentially no overhead or just the time for the MPI middleware to detect the
failure and start a new process to recover from a failure.

\subsection{Fault-tolerant QR factorization of 2D matrices}
\label{sec:caqr:ftcaqr}

TSQR is a basic block of the QR factorization. It is sufficient for tall and
skinny matrices, but achieving fault-tolerance in general matrices requires to be
also able to tolerate failures in the trailing matrix. The purpose of this paper
is to present how it can be achieved in order to implement a fault-tolerant QR
factorization for 2D, general matrices.

As stated in Section \ref{sec:caqr:algo}, the update of the trailing matrix is
made by applying it the transpose of the current panel's Q factor. If we denote
the current matrix after the factorization of the first panel as follows:
  \begin{center}
  $\begin{pmatrix} R_{0} & C_{0}' \\
    R_{1} & C_{1}' \end{pmatrix}
=  \begin{pmatrix} QR & C_{0}' \\
    ~ & C_{0}' \end{pmatrix}$
  \end{center}
The update consists of computing the $\hat{C}_{i}'$ factors on the right side of
the panel :
\begin{center}
$A = Q \begin{pmatrix} R & \hat{C}_{0}' \\
    ~ & \hat{C}_{1}' \end{pmatrix}$
  \end{center}

The blocs of the left side of the matrix are decomposed into two parts: the top
part contains as many lines as the number of columns of each block, the bottom
part contains the rest of the lines. If the width of a block is denoted by $N$
and $C[:N-1]$ denotes the first $N$ lines of matrix $C$:
\begin{center}
$C_i = \begin{pmatrix}C_{i}' \\ C_{i}''\end{pmatrix}
= \begin{pmatrix} C_{i}[:N-1] \\ C_{i}[N:] \end{pmatrix}$
\end{center}

The compact representation of the matrix is computed, as stated in section
\ref{sec:caqr:algo}, as follows:

\begin{center}
$\begin{pmatrix}\hat{C}_{0}' \\ \hat{C}_1\end{pmatrix}
= \Big(I - \begin{pmatrix}I\\Y_0\end{pmatrix} T^{T}
\begin{pmatrix}
I\\Y_{1}\end{pmatrix}^{T} \Big)
\begin{pmatrix}
C_{0}' \\ C_{1}'
\end{pmatrix}$
\end{center}

\begin{figure}[Hht]
  \resizebox{\linewidth}{!}{\input{fig_caqr_trailing}}
\caption{\label{fig:trailingupdate}Update of the trailing matrix in parallel on two processes.}
\end{figure}
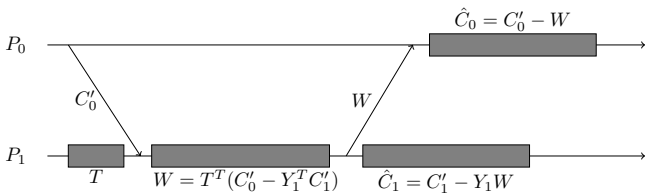

An algorithm for computing this in parallel is given in \cite{CAQR}. A graphical
representation of this algorithm in a pair of processes is given in Figure
\ref{fig:trailingupdate}, corresponding to Algorithm
\ref{algo:trailingupdate}. As noticed by \cite{CAQR}, the $T$ factors can be
computed on either process: it is on the critical path anyway.

\begin{algorithm}
  \small
    \SetKwFunction{isOdd}{isOdd}
    \SetKwFunction{mybuddy}{myBuddy}
    \SetKwFunction{send}{send}
    \SetKwFunction{recv}{recv}
    \SetKwFunction{done}{done}
    \SetKwFunction{computeT}{computeT}
    \SetKwFunction{computeY}{computeY}
    \SetKwFunction{topOfMatrix}{topOfMatrix}
    \SetKwFunction{concat}{concatenate}
    \KwData{Trailing submatrix A}
    step = 0 \;

    \While{ ! \done{} }{
      \If{ \isOdd{ step } } { \tcc{I am a sender - I am odd-numbered}
        $C_0$ = \topOfMatrix( A )\;
        $Y_0$ = \computeY() \;
        b = \mybuddy{ step }\;
        \send{ $C_0'$, b }\;
        \recv{ $W$, b }\;
        $\hat{C}_0 = C_0' - Y_0 W$\;
        \Return; \tcc{done with my part of the update}
      }
      \Else { \tcc {I am even-numbered}
        $C_1$ = \topOfMatrix( A )\;    
        $T$ = \computeT()\;
        $Y_1$ = \computeY() \;
        b = \mybuddy{ step }\;
        \recv{ $C_0'$, b }\;
        $W = T^{T}(C_{0}'+Y_{1}^{T} C_{1}')$\;
        \send{ $W$, b }\;
        $\hat{C}_1 = C_1' - Y_1 W$\;
      }
      step++\;
    }
\caption{Parallel trailing matrix update algorithm.\label{algo:trailingupdate}} 
\end{algorithm}

\begin{figure}[Hht]
  \resizebox{\linewidth}{!}{\input{fig_caqr_tree}}
\caption{\label{fig:trailingtree}Tree formed by the parallel update of the
  trailing matrix.}
\end{figure}
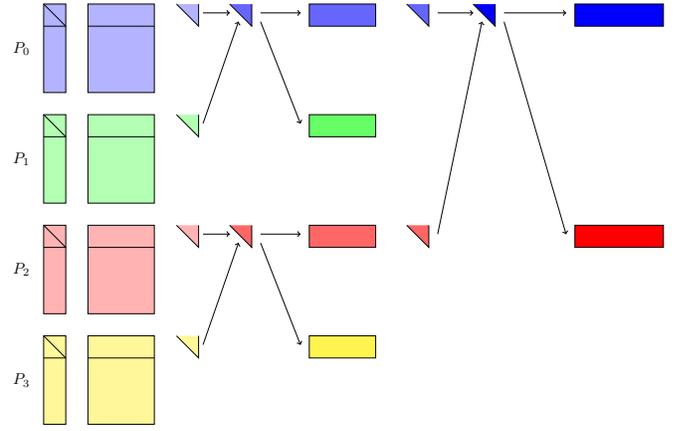

The algorithm follows a binary tree by pairs, as represented by Figure
\ref{fig:trailingtree}. We can see that, in a similar way as with TSQR, processes
exchange data and compute by pair and one of them is done with the update. As a
consequence, at each step, half of the working processes become idle.

The idea of the fault-tolerant algorithm is to use these processes that become
idle and, instead, introduce some redundancy with them. Hence, they keep
computing and the data they keep can be used to recover the state of the
computation after a process has failed and has been restarted.

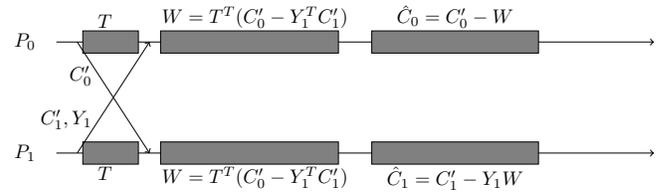
\begin{figure}[Hht]
  \resizebox{\linewidth}{!}{\input{fig_caqr_trailing_ft}}
\caption{\label{fig:trailingupdateft}Fault-tolerant update of the trailing matrix in parallel on two processes.}
\end{figure}

A graphical representation of this algorithm is given in Figure
\ref{fig:trailingupdateft} in order to give the reader the intuition behind this
algorithm. The idea is that since both processes can compute the $T$ factors,
all they need to compute their $\hat{C}_i'$ update is the other processes'
$C_j'$. With this $C_j'$, they can compute the $W$ and then their own
$\hat{C}_i'$. 

The algorithm itself is given by Algorithm \ref{algo:trailingupdateft}. We can
see that, instead of having two one-way communications in each direction between
the two processes, we have an exchange. Implemented on dual-channel
communication hardware, the latter is faster than the former, because the two
communications made by the exchange overlap. Besides, it does not increase the
length of the critical path. On the other hand, this algorithm requires both
processes to compute while of of them could be idle: it is less
energy-efficient. 

\begin{algorithm}
  \small
    \SetKwFunction{isOdd}{isOdd}
    \SetKwFunction{mybuddy}{myBuddy}
    \SetKwFunction{sendrecv}{sendrecv}
    \SetKwFunction{done}{done}
    \SetKwFunction{computeT}{computeT}
    \SetKwFunction{computeY}{computeY}
    \SetKwFunction{topOfMatrix}{topOfMatrix}
    \SetKwFunction{concat}{concatenate}
    \KwData{Trailing submatrix A}
    step = 0 \;

    \While{ ! \done{} }{
      \If{ \isOdd{ step } } { \tcc{I am a sender - I am odd-numbered}
        $C_0$ = \topOfMatrix( A )\;
        $T$ = \computeT()\;
        $Y_0$ = \computeY() \;
        b = \mybuddy{ step }\;
        \sendrecv{ $C_0'$, $C_1'+Y_1$, b }\;
        $W = T^{T}(C_{0}'+Y_{1}^{T} C_{1}')$\;
        $\hat{C}_0 = C_0' - Y_0 W$\;
        \Return; \tcc{done with my part of the update}
      }
      \Else { \tcc {I am even-numbered}
        $C_1$ = \topOfMatrix( A )\;    
        $T$ = \computeT()\;
        $Y_1$ = \computeY() \;
        b = \mybuddy{ step }\;
        \sendrecv{ $C_1'+Y_1$, $C_0'$, b }\;
        $W = T^{T}(C_{0}'+Y_{1}^{T} C_{1}')$\;
        \send{ $W$, b }\;
        $\hat{C}_1 = C_1' - Y_1 W$\;
      }
      step++\;
    }
\caption{Fault-tolerant parallel trailing matrix update algorithm.\label{algo:trailingupdateft}} 
\end{algorithm}

At the end of the execution of each step, between processes $i$ and $j$:
\begin{itemize}
\item $P_i$ has $W$, $T$, $C_i'$, $C_j'$ and $\hat{C}_i'$; therefore, if $P_j$
  fails, $P_i$ can provide the required data to recalculate $\hat{C}_j' = C_j' -
  Y_j W$ on $P_j$ (or any process that has $Y_j$)
\item $P_j$ has $W$, $T$, $C_j'$, $C_i'$, $Y_i$ and $\hat{C}_j'$; therefore, if $P_i$
  fails, $P_j$ can recalculate $\hat{C}_i' = C_i' - Y_i  W$ on $P_i$ (or any
  process that has $Y_i$) 
\end{itemize}

Therefore, the state of a failed process can be recovered using its subpart of
the initial matrix and some data kept by (at least) one process. However,
although several processes may have this data, retrieving from only one of them
is necessary. 

One minor modification would require that, instead of having $P_i$ sending $C_i'$
and $P_j$ sending $C_j'$ and $Y_j$, they both exchnge their $C_x'$ and $Y_x'$:
hence, the reconstruction would be symmetric.

\bibliographystyle{alpha}
\bibliography{ft}

\end{document}

%% file: fig_panel.tex
\begin{tikzpicture}

\draw (0,0) rectangle (4,-4 );
\draw (1,-1) rectangle (4,-4);
\node at (2, -.5) {$R$};
\node at (.5, -2) {$Q$};
\draw (0,0) -- (1, -1);
\draw [dashed] (1, -1) -- (2, -2);
\draw (2, -4) -- (2, -1);
\node [rotate=90] at ( 1.5, -2.5 ) {panel};
\node [text width=2] at ( 2.4, -2.5 ) {trailing \\matrix};

\end{tikzpicture}

%% file: fig_caqr_trailing.tex
\begin{tikzpicture}

\node [label=left:{$P_0$}] (A) at (0, 0) {};
\node (B) at (11, 0) {};
\draw [->] (A) -- (B);

\node [label=left:{$P_1$}] (C) at (0, -2) {};
\node (D) at (11, -2) {};
\draw [->] (C) -- (D);

\draw [fill=gray] ( .5, -2.2 ) rectangle ( 1.5, -1.8 );
\node [label=below:{$T$}] (E) at (1, -2) {};

\draw[->] (.5, 0) -- (1.8, -2) node[midway, left] {$C_0'$};

\draw [fill=gray] ( 2, -2.2 ) rectangle ( 5.2, -1.8 );
\node [label=below:{$W=T^{T}(C_0'-Y_1^TC_1')$}] (F) at (3.7, -2) {};

\draw[->] (5.5, -2) -- (6.7, 0) node[midway, left] {$W$};

\draw [fill=gray] ( 5.8, -2.2 ) rectangle ( 8.8, -1.8 );
\node [label=below:{$\hat{C}_1 = C_1' -  Y_1 W$}] (G) at (7.3, -2) {};
\draw [fill=gray] ( 7, .2 ) rectangle ( 10, -.2 );
\node [label=above:{$\hat{C}_0 = C_0' - W$}] (H) at (8.5, 0) {};
\end{tikzpicture}

%% file: fig_caqr_tree.tex
\begin{tikzpicture}

\draw [fill=blue!30] (0,0) rectangle (.5, -2 ) node {};
\draw (0,-.5) -- (.5, -.5 );
\draw (0,0) -- (.5, -.5 );
\node at (-.5, -1 ) {$P_0$};

\draw [fill=blue!30] (1,0) rectangle (2.5, -2 ) node {};
\draw (1,-.5) -- (2.5, -.5 );

\draw [fill=green!30] (0,-2.5) rectangle (.5, -4.5 );
\draw (0,-3) -- (.5, -3 );
\draw (0,-2.5) -- (.5, -3 );
\node at (-.5, -3.5 ) {$P_1$};

\draw [fill=green!30] (1,-2.5) rectangle (2.5, -4.5 ) node {};
\draw (1,-3) -- (2.5, -3 );

\draw [fill=red!30] (0,-5) rectangle (.5, -7 ) node {};
\draw (0,-5.5) -- (.5, -5.5 );
\draw (0,-5) -- (.5, -5.5 );
\node at (-.5, -6 ) {$P_2$};

\draw [fill=red!30] (1,-5) rectangle (2.5, -7 ) node {};
\draw (1,-5.5) -- (2.5, -5.5 );

\draw [fill=yellow!50] (0,-7.5) rectangle (.5, -9.5 );
\draw (0,-8) -- (.5, -8 );
\draw (0,-7.5) -- (.5, -8 );
\node at (-.5, -8.5 ) {$P_3$};

\draw [fill=yellow!50] (1,-7.5) rectangle (2.5, -9.5 ) node {};
\draw (1,-8) -- (2.5, -8 );

\draw [fill=blue!30] (3,0) -- (3.5, -.5 ) -- (3.5, 0);
\draw [fill=green!30] (3,-2.5) -- (3.5, -3 ) -- (3.5, -2.5);
\draw [fill=red!30] (3,-5) -- (3.5, -5.5 ) -- (3.5, -5);
\draw [fill=yellow!50] (3,-7.5) -- (3.5, -8 ) -- (3.5, -7.5);

\draw [fill=blue!60] (4.2,0) -- (4.7, -.5 ) -- (4.7, 0);
\draw [->] (3.6, -.2 ) -- (4.2, -.2);
\draw [->] (3.6, -2.7 ) -- (4.4, -.4);
\draw [->] (4.9, -.2 ) -- (5.8, -.2);
\draw [->] (4.9, -.4 ) -- (5.8, -2.7);

\draw [fill=red!60] (4.2,-5) -- (4.7, -5.5 ) -- (4.7, -5);
\draw [->] (3.6, -5.2 ) -- (4.2, -5.2);
\draw [->] (3.6, -7.7 ) -- (4.4, -5.4);
\draw [->] (4.9, -5.2 ) -- (5.8, -5.2);
\draw [->] (4.9, -5.4 ) -- (5.8, -7.7);

\draw [fill=blue!60] (6,0) rectangle (7.5, -.5 ) node {};
\draw [fill=green!60] (6,-2.5) rectangle (7.5, -3 ) node {};
\draw [fill=red!60] (6,-5) rectangle (7.5, -5.5 ) node {};
\draw [fill=yellow!80] (6,-7.5) rectangle (7.5, -8 ) node {};

\draw [fill=blue!60] (8.2,0) -- (8.7, -.5 ) -- (8.7, 0);
\draw [fill=red!60] (8.2,-5) -- (8.7, -5.5 ) -- (8.7, -5);

\draw [fill=blue] (9.7,0) -- (10.2, -.5 ) -- (10.2, 0);
\draw [->] (8.9, -.2 ) -- (9.7, -.2);
\draw [->] (8.9, -5.2 ) -- (9.9, -.4);
\draw [->] (10.4, -.2 ) -- (11.8, -.2);
\draw [->] (10.4, -.4 ) -- (11.8, -5.2);

\draw[fill=blue] (12,-.5) rectangle (14, -0 ) node {};
\draw[fill=red] (12,-5.5) rectangle (14, -5 ) node {};

\end{tikzpicture}

%% file: fig_caqr_trailing_ft.tex
\begin{tikzpicture}

\node [label=left:{$P_0$}] (A) at (0, 0) {};
\node (B) at (11, 0) {};
\draw [->] (A) -- (B);

\node [label=left:{$P_1$}] (C) at (0, -2) {};
\node (D) at (11, -2) {};
\draw [->] (C) -- (D);

\draw [fill=gray] ( .6, -2.2 ) rectangle ( 1.6, -1.8 );
\node [label=below:{$T$}] (E) at (1, -2) {};
\draw [fill=gray] ( .6, -.2 ) rectangle ( 1.6, .2 );
\node [label=above:{$T$}] (Ep) at (1, 0) {};

\draw[->] (.5, 0) -- (1.8, -2) node[pos=.3, left] {$C_0'$};
\draw[->] (.5, -2) -- (1.8, 0) node[pos=.3, left] {$C_1', Y_1$};

\draw [fill=gray] ( 2, -2.2 ) rectangle ( 5.2, -1.8 );
\node [label=below:{$W=T^{T}(C_0'-Y_1^TC_1')$}] (F) at (3.7, -2) {};

\draw [fill=gray] ( 2, .2 ) rectangle ( 5.2, -.2 );
\node [label=above:{$W=T^{T}(C_0'-Y_1^TC_1')$}] (F) at (3.7, 0) {};

\draw [fill=gray] ( 5.8, -2.2 ) rectangle ( 8.8, -1.8 );
\node [label=below:{$\hat{C}_1 = C_1' - Y_1 W$}] (G) at (7.3, -2) {};

\draw [fill=gray] ( 5.8, .2 ) rectangle ( 8.8, -.2 );
\node [label=above:{$\hat{C}_0 = C_0' - W$}] (H) at (7.3, 0) {};
\end{tikzpicture}